\documentclass[]{aa}
\usepackage{graphicx}
\usepackage{natbib}
\bibpunct{(}{)}{;}{a}{}{,} 
\usepackage{color}
\newcommand{\beq}{\begin{equation}}
\newcommand{\eeq}{\end{equation}}
\newcommand{\beqa}{\begin{eqnarray}}
\newcommand{\eeqa}{\end{eqnarray}}
\begin{document}
\title{A failed filament eruption inside a coronal mass ejection in active region 11121}



\author
{D. Kuridze$^{1,4}$, M. Mathioudakis$^{1}$, A. F Kowalski$^2$, P. H. Keys$^{1}$, D. B. Jess${^1}$, K. S. Balasubramaniam$^3$, F. P. Keenan${^1}$}
\institute{Astrophysics Research Centre, School of Mathematics and Physics, Queen's University, Belfast, BT7~1NN, Northern Ireland, UK \\
\ e-mail: dkuridze01@qub.ac.uk
\and Department of Astronomy, University of Washington, Box 351580, Seattle, WA 98195, USA
\and Air Force Research Laboratory, Solar and Solar Disturbances, Sunspot, NM 88349, USA
\and Abastumani Astrophysical Observatory, Ilia State University, G. Tsereteli 3, 0612, Tbilisi, Georgia}
\date{received / accepted }

\abstract {}
{We study the formation and evolution of a 
failed filament eruption observed in NOAA active region 11121 near the southeast limb on November 6, 2010. }
{We use a time series of SDO/AIA 304, 171, 131, 193, 335, 94  {\AA} images, SDO/HMI magnetograms, plus ROSA and ISOON H$\alpha$ images, to study the erupting active region.}
{We identify  coronal loop arcades  associated with a quadrupolar magnetic configuration, and 
show that the expansion and cancelation of the central loop arcade system over the filament 
is followed by the eruption of the filament.
The erupting filament reveals a clear helical twist and develops a same sign of writhe in the form of inverse $\gamma$-shape.}
{The observations support the ``magnetic breakout"  process with the eruption been  triggered by quadrupolar reconnection in the corona. 
We suggest that the formation mechanism of the inverse $\gamma$-shape flux rope may be the MHD helical kink instability.
The eruption has failed due to the large-scale, closed, overlying magnetic loop arcade that encloses the active region.}


\titlerunning{Multiwavelength observations of failed filament}
\authorrunning {Kuridze et al.}
\maketitle

\section{Introduction}

Solar eruptions are explosive ejections of large amounts of  plasma from the lower to the upper layers of 
the solar atmosphere and are some of the most interesting events occurring in the active Sun. 
Eruptions associated with coronal mass ejections (CMEs) are known as ``full eruption'', while those that do not lead to a CME are termed as ``failed eruptions''. 
It is generally accepted  that magnetic reconnection plays a crucial role in the process. However, the exact mechanism that drives solar eruptions remains to be identified.   
The classic ``tether cutting''  eruption  model  is based on a single, highly sheared magnetic bipole.  
This model assumes that the reconnection, which occurs deep in the sheared core region below the filament, 
removes stabilising restraints (tethers) leading the flux rope to erupt \citep{moore1,sturr1,moore2,moore3}.
Another eruption mechanism is that of the ``magnetic breakout'' model, originally proposed by \cite{ant} and \cite{ant1}. 
This requires a multipolar magnetic configuration with a central magnetic 
arcade which is the main restraining magnetic flux of the filament located underneath. 
The energy supply for the breakout eruption comes from the free magnetic energy of the filament, which 
can be efficiently stored in sheared and/or twisted non-potential magnetic configurations \citep{pri1}.
As the central arcade expands upward and reconnects with the outer antiparallel field overarching the whole region, 
it is removed and sidelobe  loops of the quadrupole are created. 
The removal of the central loop arcade, the main constraint over the filament, leads to the explosive eruption of the filament.
Both the tether cutting and breakout models describe the disruption in the  balance of the  upward-directed force 
of magnetic pressure and the downward-directed force of magnetic tension.

The MHD helical kink instability of a  magnetic flux rope anchored in the photosphere is considered  as an alternative triggering mechanism for 
solar eruptive phenomena. This instability is the  process which transforms twist (a measure of the windings of field lines about the axis of the flux rope)  
into writhe (a measure of the winding of the flux rope axis itself),  \citep{rust1}.
It occurs when the twist  exceeds a certain critical value. 
The conservation of helicity in ideal MHD requires that the resulting writhe, which can have the form of inverse $\gamma$ shape, should be the same sign as the transformed twist 
\citep{hood1,bat1,ger1,fan1,tor1, rust1, sr1}.

Several observational studies of filament eruptions presented in recent years   
appear to agree with the breakout initiation scenario \citep{aul1, ster2, man1, ji1,  gary1, deng1, poh1, al1, josh1, shen1}.   
The observations also show that reconnection, breakout and tether cutting can often all be present in the eruption
\citep{ster1}.  A combination of the magnetic breakout scenario and kink instability could be responsible for the eruption event presented by  \cite{wil1}. 

In this paper, we undertake an observational study of a "failed filament'' eruption associated with an M5.4-class flare \citep{schr,woods}.
We describe the morphology and dynamic of the loop arcades and filament before and after the eruption,
and interpret our observations in terms of the magnetic breakout and kink instability models.

 \section{Instruments and data}

\begin{figure}[t]
\begin{center}
\includegraphics[width=8.5cm]{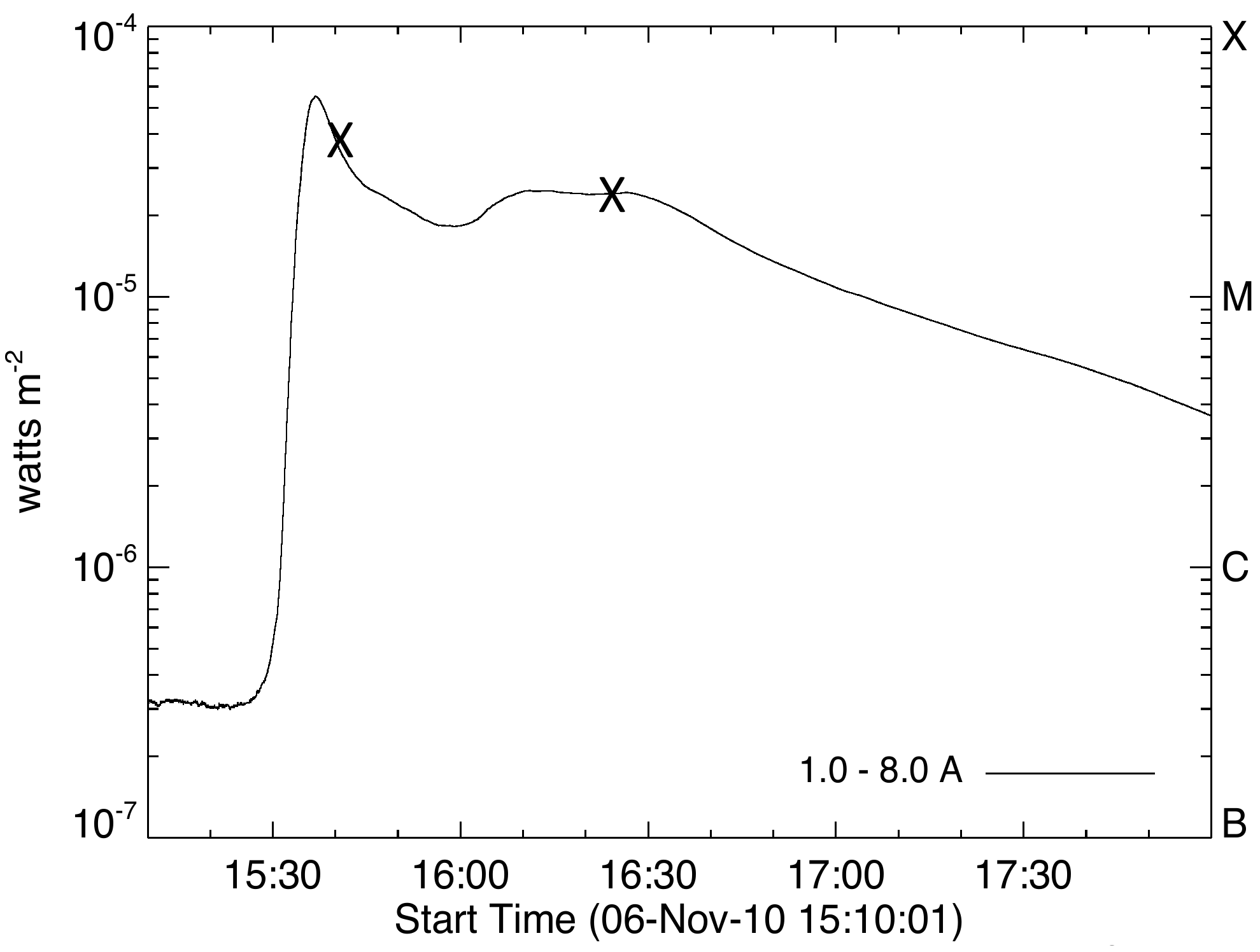}
\end{center}
\caption{GOES X-ray lightcurve (1.0 - 8.0 \AA) of the M5.4 class flare of 6 November 2010 in NOAA 11121.
The  "x"  symbols indicate the time of the filament eruptions.}
\label{fig1}
\end{figure}

\begin{figure*}[t]
\begin{center}
\includegraphics[width=18.2cm]{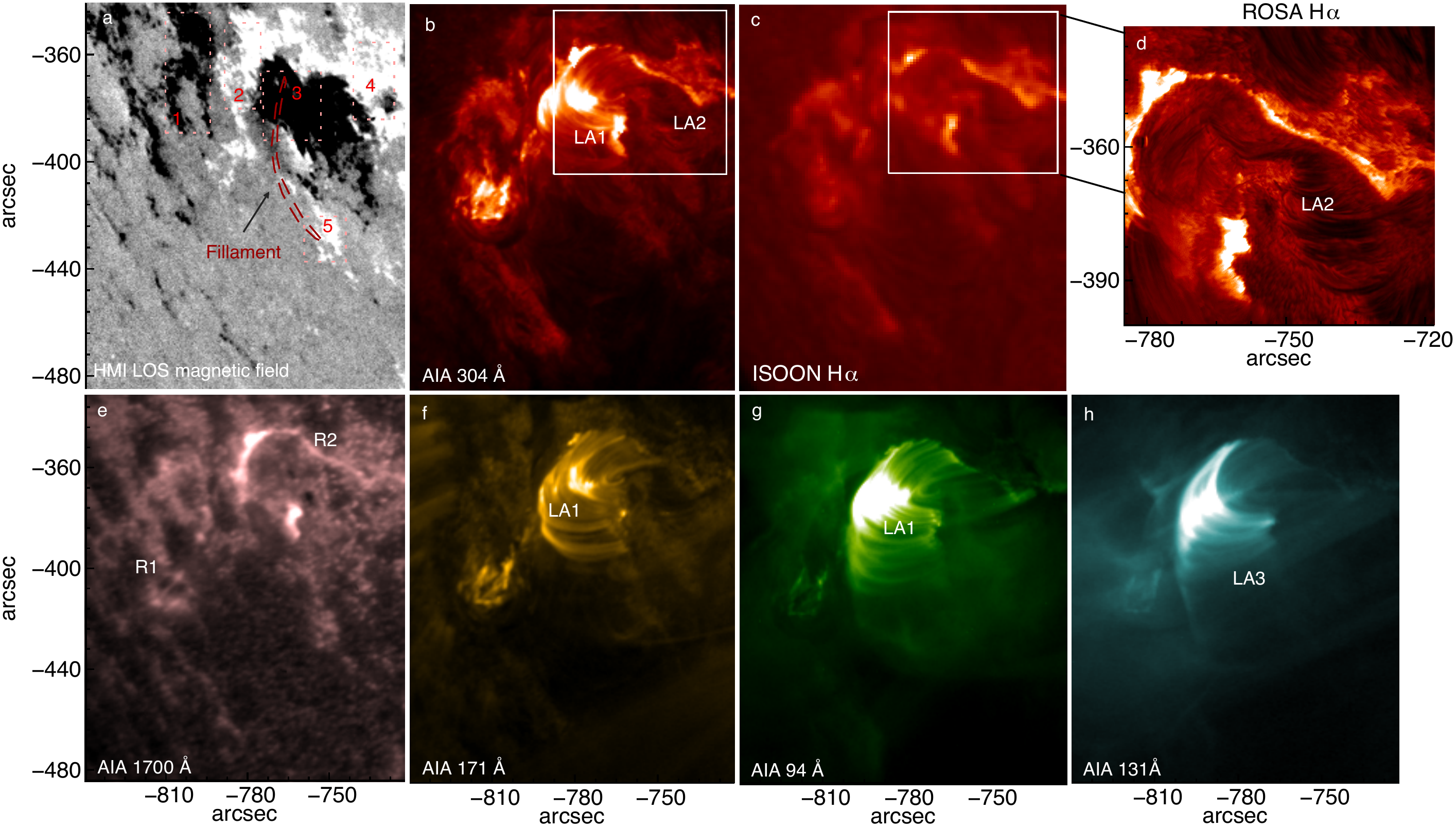}
\end{center}
\caption{Simultaneous SDO/AIA, ROSA, ISOON images and SDO/HMI magnetogram of the observed erupting region. 
Five pink, dotted boxes in the HMI  line-of-sight (LOS) magnetogram denote basic patches of positive and negative magnetic flux concentrations. The erupting filament, which is supposed to be originally located 
along the magnetic neutral line somewhere between regions 3 and 5, is indicated by the dashed lines in panel $a$. 
The large white boxes in the AIA 304 {\AA} and ISOON images indicate the ROSA H$\alpha$ field-of-view. The AIA 1700 {\AA} image shows the position of  two flare ribbons (marked as R1 and R2). 
LA1, LA2 and LA3  are three different loop arcade systems.  The temporal evolution of the AIA/ISOON/ROSA data depicted in panels $b-d$ and $f-h$ is shown in the movie provided with the online version.}
\label{fig2}
\end{figure*}

The observations were obtained between 15:10 - 18:01 UT on November 6, 2010 with the Solar Dynamics Observatory \citep[SDO;][]{lem}, the Rapid Oscillations 
in the Solar Atmosphere \citep[ROSA;][]{jess1} imaging system mounted on the Dunn Solar Telescope (DST), and
the Improved Solar Observing Optical Network (ISOON)  patrol telescope \citep{neid1, bala1}. 
SDO Atmospheric Imaging Assembly (AIA)  images  were taken in the 304, 171, 131, 91  {\AA} bandpasses  using a spatial sampling of  
0.6$''$/pixel and a 12~s cadence, with the 1700 {\AA} bandpass having a  24~s cadence. 
Due to SDO's onboard exposure-time compensation during flare activity, the AIA image sequences consisted of frames captured 
with differing exposure times (0.1--2.9~s). To compensate for this, each AIA frame was normalised to its respective exposure time, resulting in 
a time series where true changes in intensity could clearly be observed. 
The magnetic topology  of the observed active region was studied with SDO Helioseismic and Magnetic Imager \citep[HMI;][]{sch1} 
line-of-sight magnetograms. 

ROSA carried out simultaneous H$\alpha$ line core imaging over a $48''\times52''$ field-of-view between 15:59-17:01 UT. A spatial sampling of $0.069''$/pixel was used 
to match the telescope diffraction limit in the blue part of the spectrum to that of the CCD. 
In order to keep the field-of-view the same this resulted in the H$\alpha$ to be slightly oversampled 
corresponding to a spatial resolution of  $\mathrm{150~km}$ (i.e. approximately 3 pixels). High-order adaptive optics were in operation throughout the observations 
and the imaging quality was further improved using the image reconstruction algorithms of  \cite{wog1}. The effective cadence after reconstruction was 5.28~s. 
 
Full-disk H$\alpha$  line centre images were also obtained  using the ISOON  patrol telescope 
with a cadence of 1 minute and a sampling of 1.1$''$/pixel.
ISOON is operated by the USAF at the National Solar Observatory in Sunspot, NM and uses an effective
15-cm aperture telescope to image the Sun.  The images are acquired in a narrow 80 m{\AA}
bandpass in the H$\alpha$ line core, $\pm$ 0.4 A either side of line core and  in the continuum at 6300.3{\AA}. 
A 2048$\times$2048 CCD camera (14-micron pixels;  1.1$''$ sampling) obtains
images with exposure times of around 10-12 milliseconds. These are corrected for
dark-current fluctuations and compensated for changes in
light sensitivity using flat-fields acquired  immediately after the data acquisition.

\section{Observations}

The GOES light curve of the flaring active region NOAA 11121 shows that the  main flare peak occurred at around 15:37 UT (Fig.~\ref{fig1}).
A slow CME from this active region was observed by the SOHO/LASCO coronagraph at around 15:40 UT. 
The AIA and ISOON instruments did not detect this CME. However,   
the first filament eruption observed in AIA 304 {\AA} and ISOON H$\alpha$   
near the East ribbon (see panel $e$ of  Fig.~\ref{fig2}),  coincides with the CME (see online movie).
An examination of this filament does not reveal clear rising motions or extensions in its morphology.  
This suggests that the filament eruption  was   either very weak, failed/remnant part of the CME, or  it was directed in a vertical plane towards the observer making a detailed study very difficult.

The second filament eruption became evident  at around 16:24 UT during the secondary peak seen in the GOES lightcurve (Fig.~\ref{fig1}) and is the main focus of this paper.
In Fig.~\ref{fig2} we show the AIA, HMI, ROSA and ISOON images of the observed active region 
 at 15:59 UT (22 minutes after the peak of the M5.4 flare and 25 minutes before the second filament eruption). 
The HMI  line-of-sight  magnetograms  infer a multipolar magnetic field distribution. 
In the AIA  304, 171, 94 {\AA} images we can identify a loop arcade, labeled LA1, connecting  magnetic regions 2 and 3 (Fig.~\ref{fig2}).
The high spatial resolution ROSA H$\alpha$  and AIA 304, 171 {\AA} images  reveal a second  loop arcade (LA2) connecting magnetic regions 3 and 4 (see Fig.~\ref{fig2}). 
Furthermore, the AIA 131 {\AA} bandpasses show a third larger loop arcade  (LA3)  connecting magnetic regions 1 and 4 overlying the whole system.  

A comparison of the AIA 304 {\AA} images with the GOES lightcurve shows that LA1 was formed shortly after the flare peak (at around 15:37 UT)
near the main flare ribbon (R2 on AIA 1700 {\AA} panel of  Fig.~\ref{fig2} and online movie). This suggests that they are classical post-flare loop arcades. 

\begin{figure*}[t]
\begin{center}
\includegraphics[width=18.5cm]{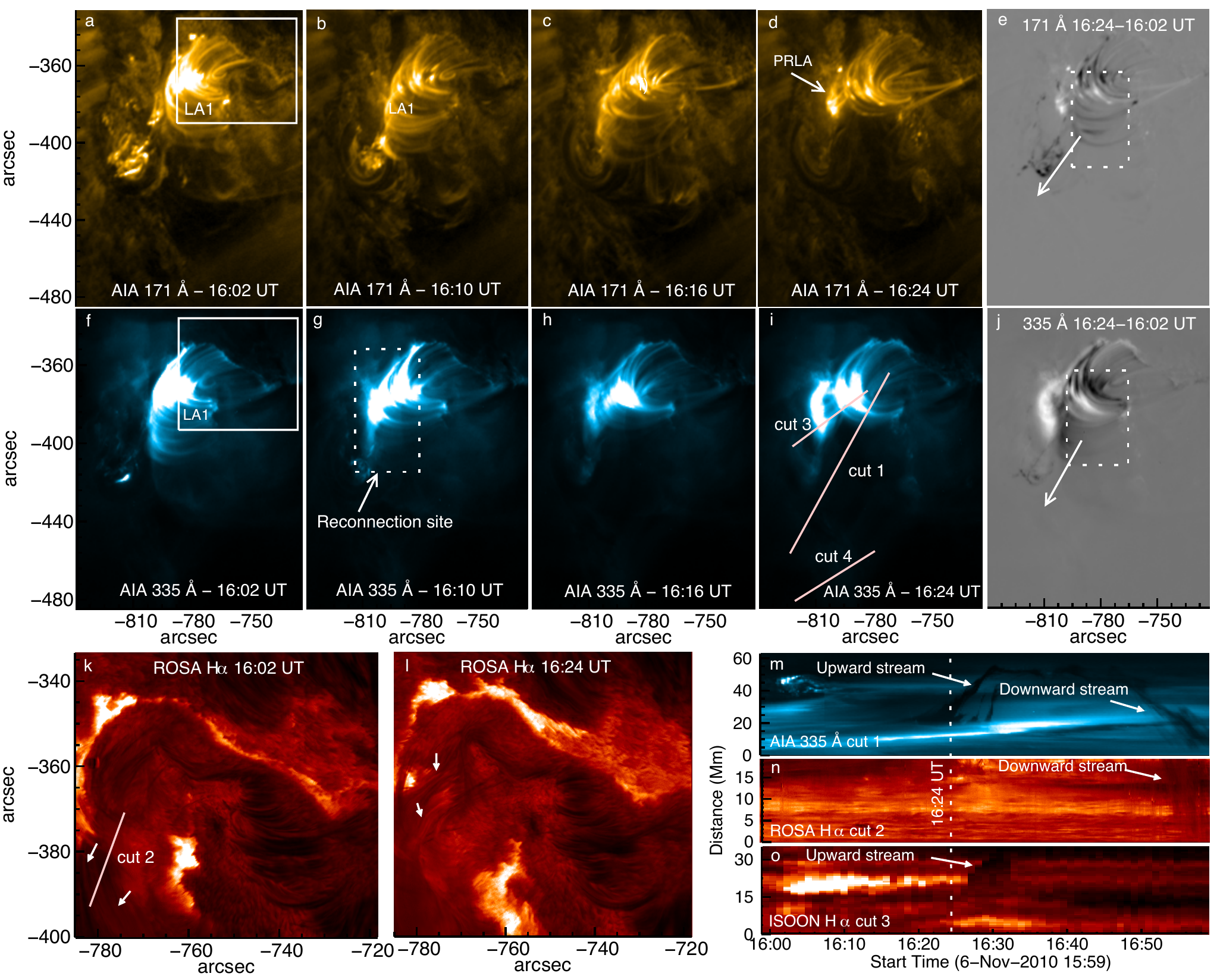}
\end{center}
\caption{Sequences of  $\mathrm{AIA~171~\AA}$ (panels $a-d$), $\mathrm{335~\AA}$  (panels $f-i$ and running difference images  (panels $e$~and~$j$)  
during the pre-eruption phase show the brightening on the top of the LA1 (referred as reconnection site), 
cancellation of LA1 and formation of the post reconnection loop arcade (PRLA).  Dotted white rectangles and 
arrows on panels $e$~and~$j$ indicate the eruptive region and the direction of the eruption. 
H$\alpha$ images (panels $k$~and~$l$) show filament threads (marked by white arrows) and their activation. 
Panels $m$~$-$~$o$ are the time-distance diagrams of the eruptive filament as observed in 335 {\AA}, ROSA  and ISOON H$\alpha$ 
lines plotted along  the $cut~1$, $2$ and $3$, respectively  in the panels $i$ and $k$. 
The vertical dashed line (panels $m$~$-$~$o$) indicates the start time of the eruption (16:24~UT). 
White arrows (panels $m~-~o$) indicate the upward (eruptions) and downward (draining)  plasma streams  passing cuts 1, 2 and 3 in panels $i$ and $k$.}
\label{fig3}
\end{figure*}

Just before the eruption, between 15:59 and 16:24 UT, the loop arcades configurations of the flare region changes  dramatically. 
Small-scale fast plasma concentrations are seen to move from the top of the LA1 downward in both directions along the loops (see online movie).
Fast jets  near the brightening on the top of LA1 are also seen to be moving in the horizontal direction (see online movie). 
A sequence of images taken at AIA 171 {\AA} and 335 {\AA} wavelengths (panels $a-d$ and $f-i$ of  Fig.~\ref{fig3}) between 16:02 and 16:24 UT
show the gradual removal of  the south-east  part of  LA1   
and the appearance of a new sidelobe arcade, labelled as PRLA  (post reconnection loop arcade)  in Figs.~\ref{fig3} and \ref{fig6} (see also online movie).
This can also be seen in the running difference images shown in panels $e$ and $j$ of Fig.~\ref{fig3},
which have been constructed  by subtracting an image at around 16:02 UT (22 minutes before eruption) from the image at 16:24 UT (just prior the eruption).
The locations of dark loops inside the white dotted boxes on panels $e$ and $j$ of Fig.~\ref{fig3}, 
and bright areas left  of the boxes, show  the removal of LA1 and appearance  of PRLA, respectively.

\begin{figure}[h]
\begin{center}
\includegraphics[width=8.97cm]{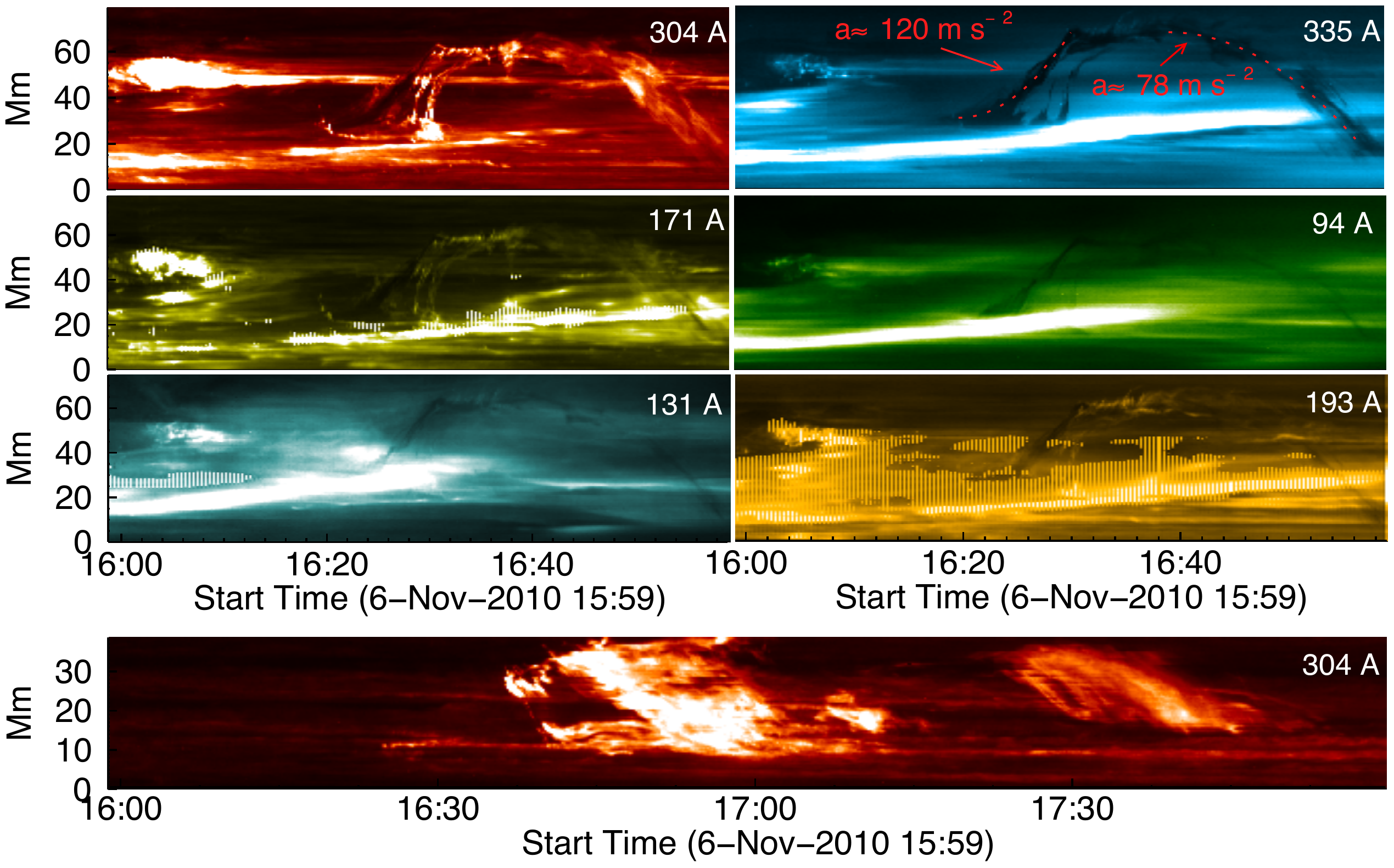}
\end{center}
\caption{Time-distance diagrams along the cut 1 (see panel $i$ of Fig.~\ref{fig3})
show the simultaneous upward motions of the eruption and downward streams in different AIA bandpasses.
The red dotted line in the top right panel is the parabolic fits to the moving features. The corresponding accelerations are also shown. 
The bottom panel is a time-distance diagram taken along the cut 4 (see panel $i$ of Fig.~\ref{fig3})  showing the downward motions of the apex plasma.}
\label{fig4}
\end{figure}

We have used a time-series of AIA 304, 171, 131, 193, 335, 94 {\AA}, ROSA and ISOON H$\alpha$ images to construct space-time diagrams ({Figs.~\ref{fig3}, \ref{fig4}). 
These show that at approximately 16:24 UT, shortly after the removal of LA1, 
the lower lying filament, located along the neutral line between the negative polarity 3 and positive polarity 5 (Fig.~\ref{fig2}),  
begins to rise and erupts (panels $m$~-~$o$ of Fig.~\ref{fig3} and online movie).
We fitted trajectories of plasma streams on the time-distance diagrams with a parabolic function (see Fig.~\ref{fig4})  and determined the average acceleration
as 120 m s$^{-2}$($\sim$0.4$g_\odot$) for the upward plasma stream. 
Its projected average speed  along cut 1 is about 65$\pm$15 km s$^{-1}$.

The eruption is shown in a sequence of snapshots in Fig.~\ref{fig5}, which also demonstrate the morphological 
changes of the erupted flux rope in the AIA 304 {\AA} and 171 {\AA} bandpasses (see also online movie). 
In the initial stages the erupted loop rose in one end while the other remained fixed (panels $b$ and $b1$ of Fig.~\ref{fig5}).  
Panels $c$, $c1$ and $e$, $e1$  of Fig.~\ref{fig5} show clear twisting motions, and a helical twist  of the erupted material 
following the formation of an inverse $\gamma$ structure (panels $d$ and $d1$ of Fig.~\ref{fig5} and online movies}).

Our multiwavelength  observations do not show evidence that during the eruption 
event some of the filament material leaves the corona and develops into a CME. 
Instead, the erupted plasma  drains  back down towards the chromosphere. 
The AIA and ROSA H$\alpha$ time-distance diagrams clearly show the downward plasma streams of the filament material (panels $m$ and $n$ of Fig.~\ref{fig3} and Fig.~\ref{fig4}) along cut 1 
with an average projected acceleration of $\sim$78 m s$^{-1}$($\sim$0.3$g_{\odot}$)  and average downward speed  of  60$\pm$10 km s$^{-1}$.
The bottom panel of Fig.~\ref{fig4} is a time-distance diagram along the $\gamma$-shape filament apex (along cut 4) showing downward motions of the apex plasma. 
Analysis of the timeseries shows that  the inverted $\gamma$-shape structure fades away and finally disappears at around 17:50 UT (see online movie). 
We note that a slow CME from this active region observed by the SOHO/LASCO coronagraph does not seem to be associated with the filament 
described in this paper and is most likely related to the first eruption occurring during the main flare peak at around 15:37 UT. 

\section{Interpretation and discussion}

\begin{figure*}[t]
\begin{center}
\includegraphics[width=18.5cm]{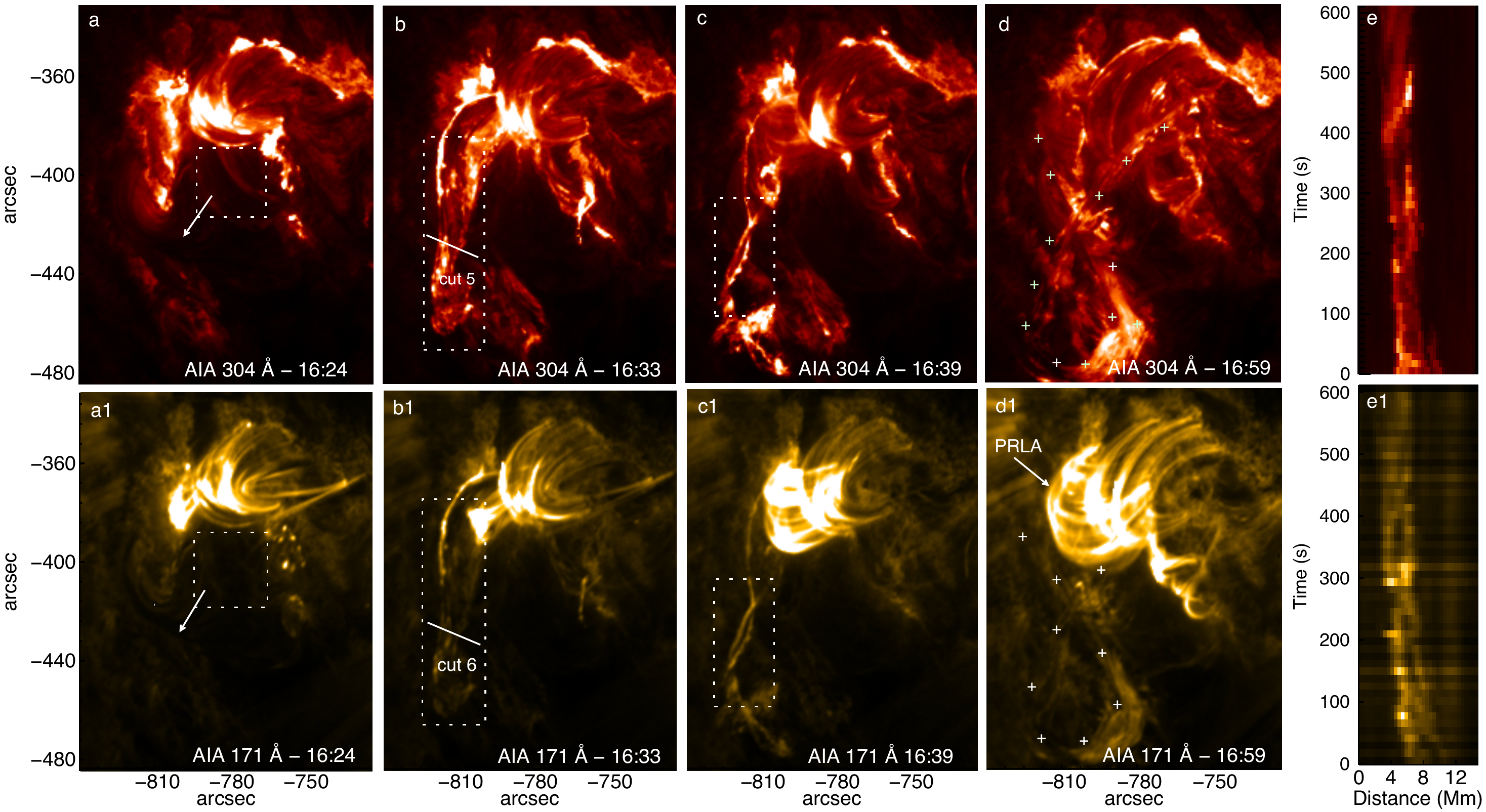}
\end{center}
\caption{ Sequences of AIA 304 and 171 {\AA} images showing the eruption and post eruption evolution of the filament. 
The erupted material shows clear  helical twist at around 16:39 UT, as seen in panels $c$ and $c1$.  
Time-distance plots along vertical cut 5 and 6 also illustrate twisting motions of the erupted flux ropes (panel $e$ and $e1$). 
The ''+'' symbols in panels $d$ and $d1$ track the inverse $\gamma$-shape structure  with apparent crossing of the two filament legs.}
\label{fig5}
\end{figure*}

Our observations suggest that magnetic breakout can explain the initiation of the observed filament eruption.
The breakout model requires a multipolar magnetic configuration which exists in the region under investigation (see Figs.~\ref{fig2}).
The high spatial resolution ROSA H$\alpha$ data indicate that the filament threads located near 
the left part of the H$\alpha$ bright ribbon  (see panel $k$ and $l$ of Fig.~\ref{fig3} and online movie), may be 
sheared magnetic structures, which can store a lot of non-potential, free magnetic energy.
The observed filament threads may therefore be considered as a non-potential core field with sufficient free energy for eruption.
We suggest that the filament is restrained by the overarching central flux arcade, LA1, seen in the 304, 171, 94 and 335 {\AA}~AIA bandpasses  (Figs.~\ref{fig2},~\ref{fig3} and online movie). 
LA2 (see ROSA H$\alpha$  and AIA 304,  171 {\AA} images of Fig.~\ref{fig2}) is a sidelobe loop arcade of the quadrupole  
with LA3 (AIA 131 {\AA} image of Fig.~\ref{fig2}) overlying the whole system. 
The expansion of LA1, seen in panels $a-d$ of Fig.~\ref{fig3}, can form the current sheet between the opposite oriented polarities LA1 and LA3,  
and the breakout reconnection between the two arcades can result 
in the gradual removal of LA1. 
The downflow of plasma blobs along the AIA 304 {\AA} loops  could be explained as a condensation 
of the hot coronal plasma in the post-flare loops (LA1) due to a thermal instability, whereas the horizontal jets coming from  the brightening on the top of LA1 could be the reconnection outflows.
A post reconnection loop arcade system, labelled PRLA  in Figs.~\ref{fig3} and \ref{fig5}  begins to form.
Removal of a sufficient portion of LA1 reduces the stabilising tension force, allowing the eruption of the lower lying filament material \citep{ant1,aul1}. 
It may be noted  that  if a part of the observed post-flare loops (LA1) are shrinking downwards 
as  expected  from the standard flare model \citep{forb},
they  can reconnect  with the opposite oriented filament field lines located underneath (Figs.~\ref{fig2}). 
This may facilitate the rise of the flux rope, together with breakout reconnection between expanding part of LA1 and LA3.


The post-eruption dynamics of the filament material show evidence of the helical kink instability, a process through which twist is transformed into writhe \citep{rust1}. 
At approximately  16:39 UT, the erupted plasma developed a clear twisted flux rope structure, as seen in panels $c$ and $c1$ from Fig.~\ref{fig5} (see also online movie).
Over the next 20 minutes  the formation of an inverse $\gamma$-shape filament occurred, and the apparent crossing of the two filament legs is observed (panels $d$ and $d1$ in Fig.~\ref{fig5}).
We note that the inverse $\gamma$-shape has been associated with the presence of a so-called helical writhe, 
which is a measure of the winding of the flux rope axis.
The conservation of helicity in ideal MHD requires that the resulting writhe must be the same sign as the initial twist \citep{rust1, green1}. 
To determine the sign of the twist and writhe, we need to define which leg of the flux rope is foreground  and which is background relative to the observer.
Fig.~\ref{fig5} (panels  $d$, $d1$) shows that the filament segment which is connected to the right footpoint  
is foreground as it obscures the filament segment which is connected to the second footpoint (see also online movie). 
This means that the writhe has a right-handed positive sign. The helical twist in the  panel $c$ and $c1$ of Fig.~\ref{fig5} has the same configuration.
Thus, we conclude that writhe and twist both have the same polarity, therefore suggesting that 
the helical kink instability of a twisted magnetic flux rope is the mechanism of the formation of the inverse $\gamma$-shape flux rope.

\begin{figure*}[t]
\begin{center}
\includegraphics[width=18.5cm]{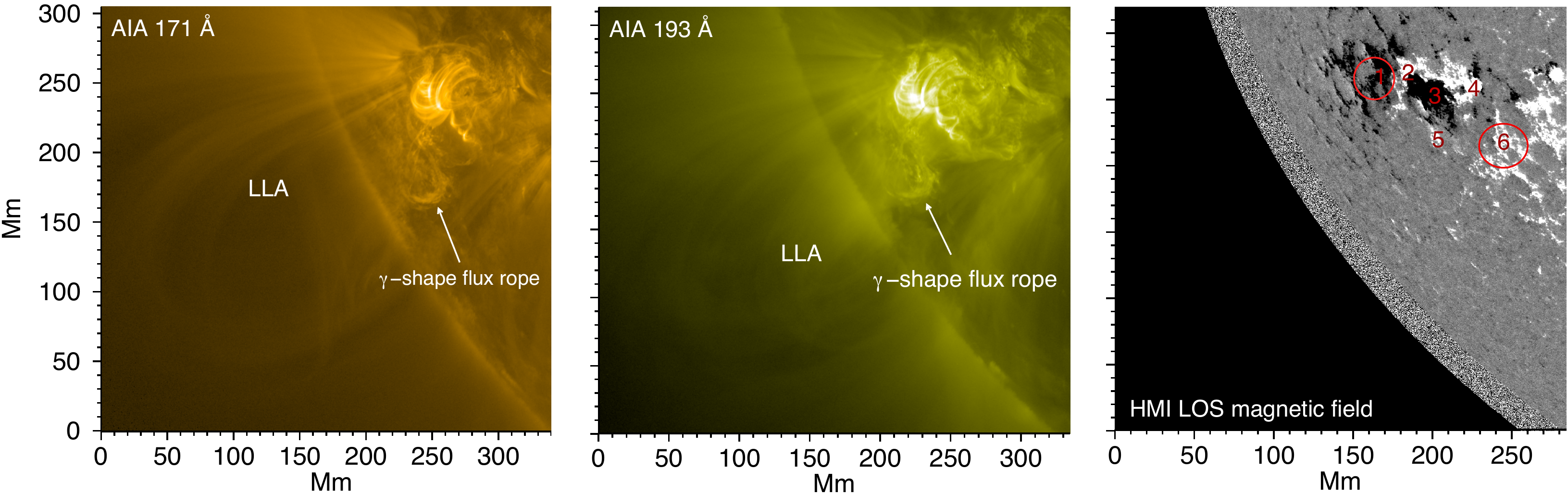}
\end{center}
\caption{AIA 171 and 193 {\AA} images showing the  large-scale, closed loops above the erupted flux rope at around 16:55 UT. 
The coresponding HMI LOS magnetogram   indicates that these loops connect magnetic polarities 1 and 6 (circled).}
\label{fig6}
\end{figure*}

\begin{figure*}[]
\begin{center}
\includegraphics[width=17.0 cm]{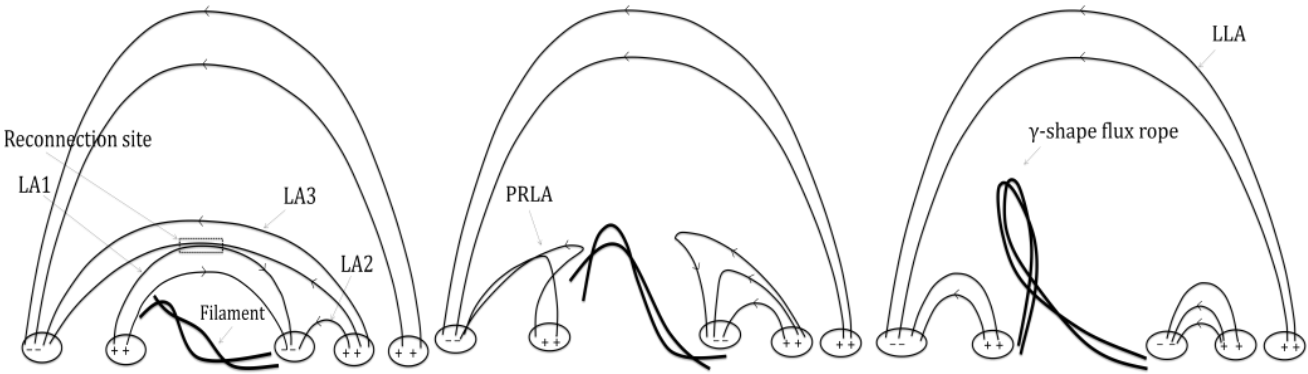}
\end{center}
\caption{A simple schematic sketch of the breakout initiation scenario for the observed  eruption, 
showing that reconnection over the filament material  between LA1 and LA3  removes the central flux system (LA1)  and results in the final eruption (left and middle panels).
The large-scale, asymmetric, closed, overlying magnetic loop arcade (LLA) could be the reason for confining the eruption (right panel).}
\label{fig7}
\end{figure*}

The eruption did not lead to a coronal mass ejection and drained back onto the solar surface.
There have been several observations of failed eruptions \citep{ji1, al1, liu1}, 
but it is still not clear exactly what defines the precise conditions that lead to a failed as opposed to a full eruption.
According to statistics, about 44~$\%$ of the eruptions associated with M-class flares do not lead to CMEs  \citep{chen1}. 
\cite{gil2} suggest that the position of the reconnection site in the inverse polarity flux rope model  can determine the outcome of an eruption, 
with reconnection below the filament most likely to produce a full eruption, 
and that above  to produce a failed eruption \citep{gil1, al1, gil2}.  
Furthermore, previous observations suggest that, 
in the magnetic breakout or kink instability scenario, the interaction between an eruptive filament and its magnetic 
environment can also play an important role in determining the nature of the eruption \citep{wil1, gibfan1, ji1, jiang1}. 
According to \cite{ji1},  if the higher-lying magnetic field lines above the filament remain close, the eruption may not propagate  into the corona 
although an overlying closed coronal magnetic field may not be a sufficient condition for a failed eruption to occur.
The simulations of \cite{tor1} showed that the kink instability could trigger a 
failed filament eruption if the overlying magnetic field decreases slowly with height.
The degree of the helical twist in the filament may also determine the nature of the eruption  \citep{rust1}. 
Other factors can include the asymmetric confinement of the background field above the filament. 
This suggests that if the filament erupts asymmetrically with respect to the overlying loop arcade, it is more likely to become a failed eruption  \citep{liu1}.
In the observations presented here, the erupted filament and the whole active region is overlaid  
by a large-scale, closed coronal loop arcade (LLA) seen in the AIA 171 and 193 {\AA} data (Fig.~\ref{fig6}). 
This coronal loop system connects magnetic polarities 1 and 6 and is not symmetric with respect to the filament (Fig.~\ref{fig6}). 
The coronal loops in LLA do not undergo any important topological changes during the whole observing period.
It  appears  that breakout reconnection has opened only the lower  magnetic  arcade (LA3) above the filament while higher  field lines remained closed (Fig.~\ref{fig6}).
This suggests that the eruption may be confined by this large-scale  loop arcade.
Unfortunately, the analysis does not reveal the precise orientation of the erupted
 $\gamma$-shape flux rope field with respect to LLA. Their orientation does not seems to favour magnetic reconnection.
A schematic diagram presented in Fig.~\ref{fig7} describes the interpretation of the observed failed filament eruption. 


 \section{Summary}

We have presented multi-instrument/multi-wavelength observations of the 
solar eruption event  in active region NOAA 11121  on 6 November  2010
using data obtained with the ROSA, SDO, and ISOON instruments. 
These show that the eruption process comprises the
pre-eruption removal of the field lines above the filament, eruption of the filament, 
development of the strong helical twist into the erupted filament, 
formation of an inverse $\gamma$-shape  structure and the draining of filament material back to the solar surface. 
A morphological study of this event supports that the magnetic breakout scenario 
and helical kink instability should be responsible for the observed evolution of the event.
The large-scale, closed, overlying magnetic loop arcade may have confined the eruption.
Future observations could focus on quantifying the amount of the helical twist, writhe and magnetic field.

\begin{acknowledgements}
We thank an anonymous referee for many important comments and suggestions especially on the interpretation part of our manuscript.
Observations were obtained at the National Solar Observatory, operated by the 
Association of Universities for Research in Astronomy, Inc (AURA) under 
agreement with the National Science Foundation. 
We thank the teams of SDO/AIA and SDO/ HMI  for providing valuable data. 
This work is supported by the Science and Technology Facilities Council (STFC).
This work has been supported by the Leverhulme Trust grant F/00203/X.
We thank the Air Force Office of Scientific Research, Air Force 
Material Command, USAF for sponsorship under grant number FA8655-09-13085.
\end{acknowledgements}


\begin{thebibliography}{}
\bibitem[Alexander et~al.(2006)]{al1}Alexander, D., Liu, R., \& Gilbert, H. R. 2006, ApJ, 653, 719
\bibitem[Antiochos(1998)]{ant}Antiochos, S. K. 1998, ApJ, 502, L181
\bibitem[Antiochos et~al.(1999)]{ant1}Antiochos, S. K., DeVore, C. R., \& Klimchuk, J. A. 1999, ApJ, 510, 485
\bibitem[Aulanier et~al.(2000)]{aul1}Aulanier, G., DeLuca, E. E., Antiochos, S. K., McMullen, R. A., \& Golub,L. 2000, ApJ, 540, 1126
\bibitem[Balasubramaniam et~al.(2010)]{bala1}Balasubramaniam, K. S., Cliver, E. W., Pevtsov, A., et al. 2010, ApJ, 723, 587
\bibitem[Baty(2001)]{bat1}Baty H. 2001, A\&A, 367, 321
\bibitem[Chen(2011)]{chen1}Chen, P. F. 2011, Liv. Rev. Sol. Phys, 8, 1
\bibitem[Deng et~al.(2005)]{deng1}Deng, N., Liu, C., Yang, G., Wang, H., \& Denker, C. 2005, ApJ, 623, 1195
\bibitem[Fan(2005)]{fan1}Fan, Y. 2005, ApJ, 630, 543
\bibitem[Forbes \& Acton(1996)]{forb}Forbes, T. G., \& Acton, L. W. 1996, ApJ, 459, 330
\bibitem[Gary \& Moore(2004)]{gary1}Gary, G. A., \& Moore, R. L. 2004, ApJ, 611, 545
\bibitem[Gerrard et~al.(2001)]{ger1}Gerrard, C. L., Arber, T. D., Hood, A. W., \& Van der Linden, R. A. M. 2001, A\&A, 373, 1089
\bibitem[Gibson \& Fan(2006)]{gibfan1}Gibson, S. E., \& Fan, Y. 2006, ApJ, 637, L65
\bibitem[Gilbert et~al.(2001)]{gil1}Gilbert, H. R., Holzer, T. E., Low, B. C., \& Burkepile, J. T. 2001, ApJ, 549,1221
\bibitem[Gilbert et~al.(2007)]{gil2}Gilbert, H., Alexander, D., \& Liu, R. 2007, Sol. Phys., 245, 287
\bibitem[Green et~al.(2007)]{green1}Green, L. M., Kliem, B., T\"or\"ok, T., van Driel-Geszotelyi, L., \& Attrill, G. D. R. 2007, Sol. Phys., 246, 365
\bibitem[Hood \& Priest(1979)]{hood1}Hood, A.W., \& Priest, E.R. 1979, Solar Phys. 64, 303.
\bibitem[Jess et al.(2010)]{jess1}Jess, D. B., Mathioudakis, M., Christian, D. J., Keenan, F. P., Ryans, R. S. I., \& Crockett, P. J. 2010, Sol. Phys., 261, 363
\bibitem[Ji et~al.(2003)]{ji1}Ji, H., Wang, H., Schmahl, E. J., Moon, Y.-J., \& Jiang, Y. 2003, ApJ, 595, L135
\bibitem[Jiang et~al.(2009)]{jiang1}Jiang, Y., Yang, J., Zheng, R., Bi, Y., \& Yang, X. 2009, ApJ, 693, 1851
\bibitem[Joshi et~al.(2007)]{josh1}Joshi, B., Manoharan, P. K., Veronig, A. M., Pant, P., \& Pandey, K. 2007, Sol. Phys., 242, 143
\bibitem[Lemen et~al.(2010)]{lem}Lemen, J. R., Title, A. M., Akin, D. J., et al. 2011, Sol. Phys., 275, 17
\bibitem[Liu et~al.(2009)]{liu1}Liu, Y., Su, J., Xu, Z., Lin, H., Shibata, K., \&  Kurokawa, H. 2009, ApJ, 696, 70
\bibitem[Manoharan \& Kundu(2003)]{man1}Manoharan, P. K., \& Kundu, M. R. 2003, ApJ, 592, 597
\bibitem[Moore \& LaBonte(1980)]{moore1}Moore, R. L., \& LaBonte, B. J. 1980, in Solar and Interplanetary
Dynamics, ed. M. Dryer \& E. Tandberg-Hanssen (Dordrecht: Reidel), 207
\bibitem[Moore \& Roumeliotis(1992)]{moore2}Moore, R. L., \& Roumeliotis, G. 1992, in Eruptive Solar Flares, ed. Z. Svestka,
B. V. Jakson, \& M. E. Machado (Berlin: Springer), 69
\bibitem[Moore et~al.(2001)]{moore3}Moore, R. L., Sterling, A. C., Hudson, H. S., \& Lemen, J. R. 2001, ApJ, 552,833
\bibitem[Neidig et~al.(1998)]{neid1}Neidig, D., et al. 1998, in ASP Conf. Ser. 140, Synoptic Solar Physics, ed. K. S.
Balasubramaniam, J. Harvey, \& D. Rabin (San Francisco: ASP), 519
\bibitem[Pohjolainen et~al.(2005)]{poh1}Pohjolainen, S., Vilmer, N., Khan, J. I., \& Hillaris, A. E. 2005, A\&A, 434, 329
\bibitem[Priest \& Forbes(2002)]{pri1}Priest, E. R., \& Forbes, T. G. 2002, A\&A Rev., 10, 313
\bibitem[Rust \& LaBonte(2005)]{rust1}Rust, D. M., \& LaBonte, B. J. 2005, ApJ, 622, L69
\bibitem[Schou et~al.(2012)]{sch1}Schou, J., Borrero, J. M., Norton, A. A., et al. 2012, Sol. Phys., 275, 327
\bibitem[Schrijver(2011)]{schr}Schrijver, C. J., 2011, 16th Cambridge Workshop on Cool Stars, Stellar Systems, and the Sun. ASP Conference Series, Vol. 448,
Edited by Christopher M. Johns-Krull, Matthew K. Browning, and Andrew A. West. San Francisco: Astronomical Society of the Pacific, 2012., p.231
\bibitem[Shen et~al.(2012)]{shen1}Shen, Y., Liu, Y., \& Su, J.  2012, ApJ, 750 
\bibitem[Srivastava et~al.(2010)]{sr1}Srivastava, A. K., Zaqarashvili, T. V., Kumar, P., \& Khodachenko, M. L. 2010, ApJ, 715, 292
\bibitem[Sterling \& Moore(2001)]{ster2}Sterling, A. C., \& Moore, R. L. 2001, ApJ, 560, 1045
\bibitem[Sterling \& Moore(2004)]{ster1}Sterling, A. C., \& Moore, R. L. 2004, ApJ, 613, 1221
\bibitem[Sturrock(1989)]{sturr1}Sturrock, P. A. 1989, Solar Phys., 121, 387
\bibitem[T\"or\"ok \& Kliem(2005)]{tor1}T\"or\"ok, T., \& Kliem, B. 2005, ApJ, 630, L97
\bibitem[Williams et~al.(2005)]{wil1}Williams, D. R., T\"or\"ok, T., D\'emoulin, P., van Driel-Gesztelyi, L., \& Kliem, B. 2005, ApJ, 628, L163
\bibitem[Woods et~al.(2011)]{woods}Woods, T. N., Hock, R., Eparvier, F., et al. 2011, ApJ, 739, 59
\bibitem[W\"{o}ger et~al.(2008)]{wog1}W\"{o}ger, F., von der L\"{u}he, O., \& Reardon, K.  2008, A\&A, 488, 375


\end{thebibliography}
\end{document}